# Control of fine particles by oscillating external forces in plasmas

Satoru Iizuka and Kazuma Sakuta

*Department of Electric Engineering, Tohoku University,*
*Sendai 980-8579, Japan*

A novel technique for the control of fine particle behavior is developed and demonstrated experimentally. The technique is called a time-averaged particle driving (TAPD) method. This method contains an application of positive pulses to two point-electrodes placed facing each other with some distance in plasmas with fine particles. When the positive pulses are applied alternatively with a repetition period that is shorter than the particle response time, the particles feel only time-averaged force because of the large mass and are transported around the center between two point-electrodes. This method is quite effective for converging fine particles in the plasma.

1. **Introduction**

Fine particles of micron size in plasmas, charged negatively, are confined under the balance between electrostatic force and gravitational force. Since Coulomb-coupling parameter $\Gamma$ is large, the particles reveal various characteristic behaviors concerned with a strongly coupled state. In our previous experiments it was demonstrated that the externally applied electric and magnetic fields drove a collective motion of particles [1,2].

In order to produce a three-dimensional Coulomb crystal the gravity gives an undesirable influence. The particles are inevitably trapped in the sheath with large potential gradient where the ions are accelerated from the plasma, giving rise to a collective effect through the Coulomb collision with the particles. For example the particles are aligned in vertical direction, therefore an asymmetric Coulomb crystal is formed in the inhomogeneous sheath region. For this reason the experiment under the microgravity condition has been proposed and characteristic feature of fine particles has been reported [3,4]. We have also clarified dynamic properties of fine particles under the microgravity condition [5].

In a parallel plate rf discharge there appeared frequently a void in the fine-particle clouds. The main reason is that the fine particles are pushed toward the plasma periphery by the ion drag force and are transported to the position where the ion drag force balances with the electrostatic force directing inward. Therefore, the elimination of the void is a crucial subject for investigating the ideal Coulomb crystal in plasmas. In this paper we propose a noble method for the control of particle position by using a time-averaged particle driving (TAPD) method [6].

2. **Experimental apparatus and methods**

The method contains very simple time-varying forces acting on the particles, i.e., electrostatic force and ion drag force caused by additional point electrodes. Because of the large mass the particles feel average force varying in time at a frequency ω/2π. Here, two small point electrodes generate such additional force to drive the particles. Finally, the particles are transported toward a fixed position in the plasma.

The experiment is carried out inside a vacuum chamber of 11.6-cm wide, 15-cm height and 11.6-cm long. Top and two side-walls contain square view windows of 8 cm × 8 cm in length for laser beam injection and for observation of right-angle scattering from the fine particles, respectively. The plasma is produced by an rf electrode of 10 cm in diameter placed at the bottom of the chamber. The rf power at 13.56 MHz is applied to the electrode for the rf plasma production. The rf power can be changed in the range up to 10 W and argon gas is introduced at fixed pressure of 20 Pa. In order to confine the particles in radial direction a plate C of 8 cm × 4 cm and 1 mm thick with a combined circular hole A and B at the center is introduced on the rf electrode D as shown in Fig. 1. The diameter of the holes is 20 mm and the distance between the hole centers is 15 mm. Therefore, the holes are overlapped and the shape looks like a pair of glasses. This plate is placed on the rf electrode directly and connected electrically with the rf electrode. Fine particles used here are acrylic sphere of 10 μm in diameter, which is dropped from the dust dispenser placed at 5 cm above the rf electrode. The dust dispenser can be movable in the radial direction, therefore it is drawn into a side port after injecting the fine particles into the plasma. The particles are levitated at a few mm above the electrode.

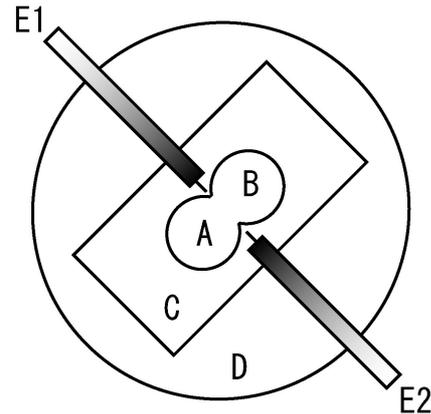

Fig. 1 Experimental apparatus. Plate C of 1 mm thick with hole A and B are placed on rf electrode D. E1 and E2 are point electrodes.

For the control of the fine particle positions two additional point electrodes E1 and E2 are introduced on the levitation electrode C as shown in Fig. 1. The tip of the point electrode consists of a wire of 0.45 mm in diameter and 3 mm long. These electrodes are facing each other above the hole A and B of the levitation electrode C in such a way that the line connecting two electrodes passes the cross points of the overlapped hole A and B as shown in Fig. 1. Low frequency time-varying voltage is applied to the tips of the electrodes alternatively with the repetition frequency of ω/2π.

## 3. Experimental results and discussions

We first investigate the effect of the point electrode biased at the potential $V_b$. When the potential $V_b$ is negative or floating potential $V_f$, the plasma potential near this point electrode is not much changed and almost uniform plasma potential is observed as shown in Fig. 2. However, with increasing the bias voltage $V_b$ the plasma potential starts being changed drastically as shown in Fig. 2. For $V_b > 37.5$ V the potential is increased near the point electrode. Since the plasma potential far away from the pint electrode is not so much changed, there appears a potential gradient near the point electrode. This suggests that an electric field is generated directing outward from the point electrode.

When the bias voltage is further increased, the plasma potential is increased in the whole region. Even in this case, there still appears a potential gradient near the point electrode. Therefore, the electric field directed outward is generated. We also observed that the increase of bias potential is accompanied by a local plasma production. We measure the spatial profile of electron density around the point electrode. For the bias voltage $V_b < 37.5$ the density profile is not changed. This is consistent with the result of potential profile. However, the electron density starts increasing locally around the point electrode for $V_b > 37.5$ V. When $V_b$ is further increased, the electron density increases in the whole region with a steep density gradient near the point electrode. Therefore, it is found that not only the plasma potential but also the plasma density is increased near the point electrode by a local discharge at the point electrode for the large positive bias voltage $V_b$.

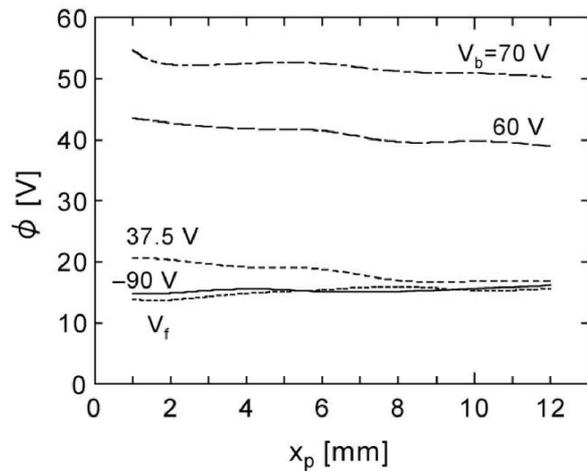

Fig. 2 Profile of plasma potential $\phi$ at position $x_p$ from the point electrode with bias potential $V_b$ as a parameter.

When the electron density around the tip of the point electrode is increased, there will appear the ion drag force directing outward from the point electrode. Then, the particles are repelled from the point electrode to reach the balance position with the electric force. In our experimental condition, however, it is found that the ion drag force is less than the electric force in the whole region.

From the results shown in Fig. 2 we find that the particles levitated above the hole of the levitation electrode can be accelerated by the electric field induced by the point electrode when it is biased positively. The negatively charged particles are accelerated toward the positively biased point electrode. If the potential of the point electrodes E1 and E2 is alternatively biased with a repetition frequency $\omega/2\pi$, the fine particles will move along a zigzag way. That is, first the particles start moving toward E1 when only E1 is turning on. But, in the next moment they turn to E2 when only E2 is turning on. For a low frequency operation the amplitude of such zigzag motion of particles is large. However, for a high frequency operation the fine particles feel only a time-averaged force and move more smoothly. They are always accelerated toward the middle point between two point electrodes. Therefore, all fine particles are gathered around this middle point wherever they are. This is an essential point of this method. We can collect fine particles in independence of their initial position. We call this technique the time-averaged particle driving (TAPD) method.

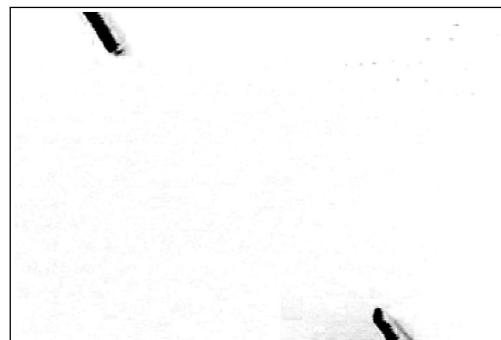

Fig. 3 Particle behaviors before applying the TAPD method. Upper left and lower right rods show the tips of point electrodes E1 and E2 shown in Fig. 1. The distance between the tips is 10 mm.

The particle response on the point electrodes is demonstrated in Figs. 3 and 4. Before applying the positive bias voltage $V_b$ to the electrodes, the particles are mainly levitated above the center region of the hole A and B, because the plasma potential near the center of the hole is relatively high. Therefore, there exist almost no particles near the point electrodes as shown in Fig. 3. On the other hand, as soon as the time varying voltage with frequency of 100 Hz is applied to both electrodes, the particles are pulled toward the center between two electrodes as shown in Fig. 4.

First, we can observe a motion of small particle cloud coming from the region B after turning on the bias voltage as shown in Fig. 4(a). Then, this particle cloud is mixed with a small particle cloud coming from the region A. Then, stationary particle cloud is formed at 3 s after turning on the voltage as shown in Fig. 4(b). The response time is of the order of 1 s depending on the applied voltage $V_b$. The bias voltage to the point electrode varies the shape of the particle cloud in the steady state. The shape is almost circular for lower voltage. However, it varies to ellipse for higher voltage. When the voltage is turned off, the particles return to the initial position and we cannot observe the particles between E1 and E2 as shown in Fig. 3. Therefore, the application of the time varying voltage to the point electrodes is quite effective for collecting fine particle clouds. When the frequency is decreased, we observe a large amplitude fluctuation of particles. The amplitude is increased with decreasing the frequency.

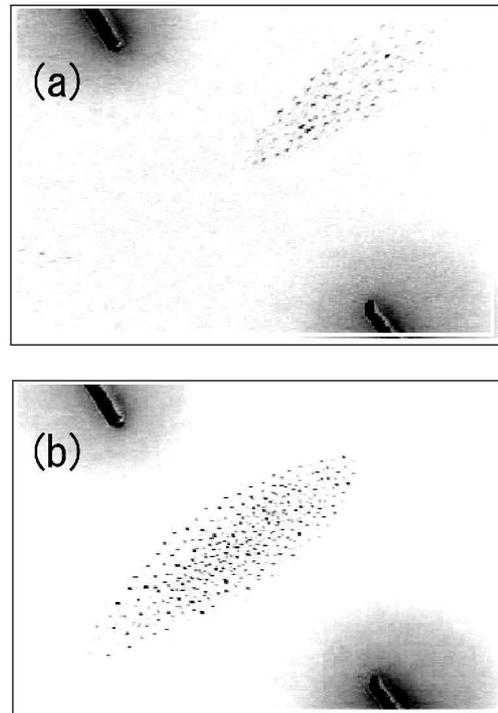

Fig. 4 Particle behaviors at (a) 2/3 s and (b) 3 s after applying the TAPD method. Upper left and lower right rods show the tips of point electrodes E1 and E2 shown in Fig. 1. The distance between the tips is 10 mm.

## 4. Conclusion

We have demonstrated the control of the particle position by using a new technique of the time-averaged particle driving (TAPD) method. This method is quite important for transporting the particles to a fixed position in plasmas. This method is also quite effective for creating a three-dimensionally symmetric Coulomb crystal in the plasma under the microgravity condition even when the void is formed in the particle cloud.

The work was supported by a Grant-in-Aid for Scientific Research from the Ministry of Education, Culture, Science and Technology, Japan.